\documentclass[a4paper,fleqn,usenatbib,referee]{mnras}

\usepackage[T1]{fontenc}
\usepackage{ae,aecompl}

\usepackage{graphicx}   
\usepackage{amsmath}    
\usepackage{amssymb}    
\usepackage{multirow}

\def \kms {{ \rm km\;s$^{-1}$}}

\graphicspath{{figs/}}

\newcommand{\oql}{O$^{7+}$/O$^{6+}$}

\newcommand{\velunit}{km~s$^{-1}$}

\newcommand{\ahe}{$A_{He}$}
\newcommand{\vap}{$v_{\alpha p}$}
\newcommand{\vratio}{$v_{\alpha p}/v_{A}$}

\title[]{Helium abundance and speed difference between helium ions and protons in the solar wind from coronal holes, active regions, and quiet Sun
}

\author[Hui Fu et al.]{Hui Fu$^{1}$, M.~S. Madjarska$^{2}$, Bo Li$^{1}$, LiDong Xia$^{1}$, ZhengHua Huang$^{1}$
\\
$^{1}$Shandong Provincial Key Laboratory of Optical Astronomy and Solar-Terrestrial Environment, Institute of Space Sciences,\\ Shandong University, Weihai, 264209 Shandong, China, fuhui@sdu.edu.cn\\
$^{2}${Max Planck Institute for Solar System Research, Justus-von-Liebig-Weg 3, 37077, G\"ottingen, Germany, madjarska@mps.mpg.de}\\
}

\pubyear{2018}

\date{Accepted XXX. Received YYY; in original form ZZZ}

\pubyear{2018}

\begin{document}
\label{firstpage}
\pagerange{\pageref{firstpage}--\pageref{lastpage}}
\maketitle

\begin{abstract}
Two main models have been developed to explain the mechanisms of
    release, heating and acceleration of the nascent solar wind,
    the wave-turbulence-driven (WTD) models and reconnection-loop-opening (RLO) models,
    in which the plasma release processes are fundamentally different.
Given that the statistical observational properties of helium ions
    produced in magnetically diverse solar regions could provide
    valuable information for the solar wind modelling, we examine the statistical properties of
    the helium abundance (\ahe)  and  the speed difference between helium ions and protons (\vap)
    for coronal holes (CHs), active regions (ARs) and the quiet Sun (QS).
We find  bimodal distributions in the space of  \ahe\ and  \vratio\
    (where $v_{A}$ is the local Alfv\'en speed)
    for the solar wind as a whole.
The CH wind measurements are concentrated at
     higher \ahe\ and \vratio\ values with a smaller \ahe\ distribution range,
    while the AR and QS wind is associated with lower \ahe\ and \vratio,
     and a larger \ahe\ distribution range.
The magnetic diversity of the source regions and the physical processes related to it
    are possibly responsible  for the different properties of \ahe\ and \vratio.
The statistical results suggest that the two solar wind generation mechanisms,
    WTD and RLO,
    work in parallel in all solar wind source regions.
In CH regions  WTD plays a major role,
    whereas the RLO mechanism is more important in AR and QS.
\end{abstract}


\section{Introduction}
\label{sect_intro}

Helium is ranked as the second most abundant element in the Sun and
    in the solar wind (SW), and it is an important tool in exploring the nature of the solar wind.
In particular, the difference in the helium ion and  proton properties
    can help us understand the mechanisms for the release, heating and acceleration
    of the nascent solar wind
    \citep[e.g.,][]{1982JGR....87...35M, 1996JGR...10117047N,1996GeoRL..23.1183S,
    2001JGR...106.5693R, 2007ApJ...660..901K, 2012ApJ...745..162K}.
The abundance of helium (\ahe) and the speed difference
    between helium ions and protons (\vap) in the solar wind
    were extensively studied in the past.
The abundance of helium is about 8.5\% in the photosphere
    \citep[e.g.,][]{1998SSRv...85..161G,2009ARA&A..47..481A}.
Measurements of the corona above polar coronal holes and surrounding quiet Sun areas showed that
 \ahe\ is in the range 4\% -- 5\%   \citep{2001ApJ...546..552L,2003ApJ...591.1257L}.
The \ahe\ is usually below 5\%\ in the solar wind and changes with the solar activity
  \citep{1974JGR....79.4595O,1978JGR....83.2177F}.
Using data obtained by WIND,
    \citet{2001GeoRL..28.2767A}
confirmed this finding and also established that this tendency is more clear for the slow SW.
By dividing the solar wind into 25 speed intervals,
    \citet{2007ApJ...660..901K,2012ApJ...745..162K}
    examined the relationship between the helium abundance
    and the speed of the solar wind for a whole solar activity cycle, and found
     a strong correlation between \ahe\ and sunspot numbers for the slowest solar wind.

The speeds of helium ions are usually larger than the proton speeds in the solar wind,
    although helium ions are heavier than protons.
Using data obtained by Helios,
    \citet{1982JGR....87...35M}
    analysed the speed difference between helium ions and protons,
    and  found that \vap\ increases with the solar wind speed.
While  \vap\ is close to the local Alfv\'en wave speed in the fast SW,
    the average \vap\ for the slow SW is close to zero,
    and  \vap\ in  the fast SW  decreases with the increase of the heliocentric distances
    at almost the same rate as of  $v_{A}$.
Consequently, these results were confirmed from  observations
   made by Ulysses
    \citep{1996JGR...10117047N,2001JGR...106.5693R},
    Wind
    \citep{1996GeoRL..23.1183S},
    and ACE
    \citep{2011PhRvL.106o1103B}.

The plasma release, heating and acceleration mechanisms of the nascent solar wind are a fundamental problem
    in solar and space physics.
 Two classes of models,
    the wave-turbulence-driven (WTD) models
    \citep{1986JGR....91.4111H,1991ApJ...372L..45W,
    2007ApJS..171..520C,2009EM&P..104..121V}
    and the reconnection loop opening (RLO) models
    \citep{1999JGR...10419765F, 2003JGRA..108.1157F, 2003ApJ...599.1395S,
    2004ApJ...612.1171W, 2006JGRA..111.9115F}
    have been proposed to account for this.
The essential difference between the two models is that
    the plasma escapes directly along open magnetic field lines
  in the WTD models,
    whereas in the RLO models the plasma is released by reconnection
    between open magnetic field lines and closed loops.
Waves and turbulence are all important in the two plasma release mechanisms.
In the RLO models waves originate in the reconnection process,
    while waves are generated by photospheric motions in the WTD models
    \citep[e.g.,][]{ 2003JGRA..108.1157F, 2007ApJS..171..520C,
            2009LRSP....6....3C, 2016SSRv..201...55A}.
Determining which physical mechanisms
    are at work and/or the extend of the contribution
    of any of the mechanisms is prerequisite for establishing physically realistic
    models of the solar wind and the heliosphere \citep{2009LRSP....6....3C}.

Generally, the solar wind is categorised by speed.
However, the speed is not the only classification criterion of the solar wind
    \citep{2012SSRv..172..169A,2016SSRv..201...55A}.
The solar wind can also be differentiated by
some of its in-situ measured properties, like for instance
    the charge state \citep{2009GeoRL..3614104Z, 2012ApJ...744..100L, 2014ApJ...793...44Z}.
As the charge state  does not change beyond several solar radii, it carries direct  information
about the temperature of the source region
    \citep{1983ApJ...275..354O,1986SoPh..103..347B}.
Based on the in-situ properties of the proton number density, proton temperature, magnetic filed strength,
    and solar wind speed,
    \citet{2015JGRA..120...70X} classified the solar wind into four categories,
    coronal-hole-origin plasma, streamer-belt-origin plasma, sector-reversal-region plasma, and ejecta.
In addition, on the basis of the above four category classification,
    \citet{2017JGRA..12210910C} developed a solar wind classification algorithm using a Machine Learning algorithm.

The solar wind can also be classified by source regions
    \citep{2002JGRA..107.1488N, 2004SoPh..223..209L, 2015SoPh..290.1399F,
            2017ApJ...836..169F, 2017ApJS..228....4Z,2017ApJ...846..135Z}.
This is reasonable
    as CHs, ARs, and QS  are all regarded as the sources of the solar wind, but their actual contribution is still uncertain. How the solar wind is produced in these regions is also still debatable.
It is generally accepted that CHs are the sources of the solar wind
    \citep[e.g.,][]{1973SoPh...29..505K,1999SSRv...89...21G}.
The solar wind can also  originate from QS regions
    \citep[e.g.,][]{2000JGR...10512667W, 2005JGRA..110.7109F,
    2015SoPh..290.1399F}.
Another source region of the solar wind that has been investigated
    in detail recent years are the edges of active regions.
From the comparison of the velocity distributions at 2.5~R$_{\odot}$
    and potential field extrapolations using Kitt Peak magnetograms,
    \citet{1999JGR...10416993K}
    found that low-speed wind regions are associated with large magnetic field expansions
    originating from area adjacent to ARs.
Shortly after \citet{2001ApJ...553L..81W} established the presence of intermittent  flows
    with velocities in the range of 5 -- 20~\kms\ at the edge of an AR
    using data from the Transition Region and Coronal Explorer (TRACE)
    in the Fe~{\sc ix/x}~171~\AA\ passband.
With the launch of the Hinode spacecraft it became possible to further investigate
    these regions using imaging and spectroscopic data.
\citet{2007Sci...318.1585S} identified the existence of outflowing plasma
    at the periphery of ARs in images taken with the X-ray Telescope (XRT)
    on board Hinode and obtained upward Doppler velocities of $\sim$50~\kms\ using the Extreme-ultraviolet Imaging Spectrometer (EIS) on Hinode in an Fe~{\sc xii} line
    (no wavelength is mentioned in the article).
Later, \citet{2008A&A...481L..49D} and \citet{2008ApJ...676L.147H} investigated
    these outflows obtaining upflow Doppler velocities ranging from 5 to 50~\kms\ in coronal lines.
\citet{2010ApJ...715.1012B} established that the outflows can be present for
    several days and also identified multiple velocity components in the
    EIS Fe~{\sc xii} and Fe~{\sc xiii}  lines of up to 200~\kms.
The follow-up finding of upflows at  heights between 1.5 and 2.5 R$_{\odot}$
    in the solar atmosphere  using data  from the Ultra-Violet Coronagraph Spectrometer
    on board SoHO in the H~{\sc i} Ly $\alpha$ and O~{\sc vi} doublet lines
    at 1031.9~\AA\ and 1037.6~\AA\  \citep{2012A&A...545A...8Z} further supported
    the evidence that AR peripheries are a possible source of the
    slow SW.
Several studies  followed  investigating both observationally and through modelling
    the association of AR upflows with the solar wind \citep{2010A&A...516A..14H, 2012SoPh..281..237V, 2014SoPh..289.3799C,2014SoPh..289.4151M, 2015NatCo...6E5947B, 2015A&A...584A..39G, 2015A&A...584A..38V, 2016A&A...594A..40Z, 2017SoPh..292...46B}.

In the present paper, we analyse  \ahe\ and \vratio\  (where $v_{A}$ is the local Alfv\'en speed) of the solar wind for
    the three general types of solar regions,
    coronal holes (CHs), active regions (ARs) and the quiet Sun (QS), during three phases of the solar activity cycle. The magnetic field structures are significantly different in these three regions, with
CHs generally occupied by large scale open magnetic field lines,
   whereas AR and QS regions are mainly taken up by closed loops
   \citep{2004SoPh..225..227W, 2010ApJ...719..131I, 2014A&ARv..22...78W}.
 We aim at demonstrating the differences  in the properties of \ahe\ and \vratio\
  for the three source region solar wind and during three phases of the solar cycle activity.
Our expectations are that  statistical observational results on
    the variabilities of  \ahe\ and \vratio\ produced in magnetically diverse solar regions
    may provide helpful and valuable information for the solar wind modelling.

The paper is organised as follows:
In Section~\ref{sec:data}, we describe the data and our method of analysis.
The statistical results are presented and discussed in Section~\ref{sec:results and discussion}.
    Section~\ref{sec:conclusion} summarises the present study,
    ending with concluding remarks.

\section{Data and Analysis} \label{sec:data}

The data used in this study are all obtained by the WIND spacecraft.
The proton and helium ion velocities and densities
    were recorded by the Solar Wind Experiment (SWE) Faraday Cup instruments
    \citep{1995SSRv...71...55O}. The \ahe\ was obtained from the density ratio between the helium ions and protons.
The magnetic field was measured by the Magnetic Field Investigation (MFI) \citep{1995SSRv...71..207L}.
The speed uncertainties of the solar wind are less than 0.16\%
    \citep{2006JGRA..111.3105K}.
In some cases, the SWE could not yield accurate helium ion measurements  \citep{1996GeoRL..23.1183S}. First, when the proton energy-per-charge distribution is very broad, it makes the helium ion signal overpowered by the proton signal, and thus the helium ion signal cannot be extracted. Second, when the helium ion flux is unusually low, it is down the detection threshold of the detectors. Third, if the solar wind speed is too high, the helium ions may exceed the highest energy-per-charge step of SWE, thus becoming undetectable.
In order to ensure accurateness,
    we only use data that are free of the above mentioned discrepancies.
The time resolution of the data is 92 seconds. The data were averaged over 1~hr.
Generally, the direction of the \vap\ is assumed along the magnetic field lines
     \citep[e.g.,][]{1976JGR....81.2719A,1982JGR....87...35M,1996GeoRL..23.1183S,
     2011PhRvL.106o1103B,2001JGR...106.5693R}.
The \vap\ was calculated as
    $v_{ap}=(v_{ra}-v_{rp})/cos(\theta)$,
    where $v_{ra}$ and $v_{rp}$ are the radial speed of the helium ions and protons, respectively,
    and $\theta$ is the angle between the radial vector and the magnetic field.
In order to reduce the uncertainties,
    we discarded the observations in which $\theta$
    is greater than 72.5 degrees
    ($cos(\theta) < 0.3$) as done in
    \citet{2001JGR...106.5693R}.
The \vap\ is usually compared with the local Alfv\'en speed
    \citep[$v_{A}$;][]{1982JGR....87...35M,1996GeoRL..23.1183S,
    2011PhRvL.106o1103B,2001JGR...106.5693R}
    which was calculated as
    $ v_{A}=22.3*B_{rtn}/\sqrt{(n_p+4*n_a)}$,
    where $B_{rtn}$, $n_p$, and $n_a$ are the magnetic filed strength in nT,
    and density of proton and helium ions in number per cubic centimetre (n/cc).
At last,  \vap\ was divided by $v_{A}$.

For completeness, we repeat here the description of  the two-step mapping procedure
    \citep{1998JGR...10314587N,2002JGRA..107.1488N,2004SoPh..223..209L}
   that was used in tracing the solar wind back to the solar surface (already described in \citet{2015SoPh..290.1399F,2017ApJ...836..169F}).
 First, each  solar wind parcel is traced back in a ballistic approach
    to the source surface which is
    implemented by the coronal magnetic field model.
Second, the wind parcel is traced from the source surface to the photosphere by
    following the magnetic filed lines computed by a potential field source surface (PFSS)
    model \citep{1969SoPh....6..442S,1969SoPh....9..131A}.
The footpoints of the solar wind parcels were then placed on the photospheric magnetograms  obtained with the Michelson Doppler imager
    \citep[MDI,][]{1995SoPh..162..129S}
    and the EUV images taken by
    the Extreme-ultraviolet Imaging Telescope    \citep[EIT,][]{1995SoPh..162..291D}  on board Solar and Heliospheric Observatory
    \citep[SoHO,][]{1995SoPh..162....1D}.
The regions with the located footpoints were then categorised into three groups.
The solar wind was named by the three type of source regions they originate from,
    CH, AR, or QS wind.

\begin{figure*}
\begin{center}
\centerline{\includegraphics[width=1.0\textwidth]{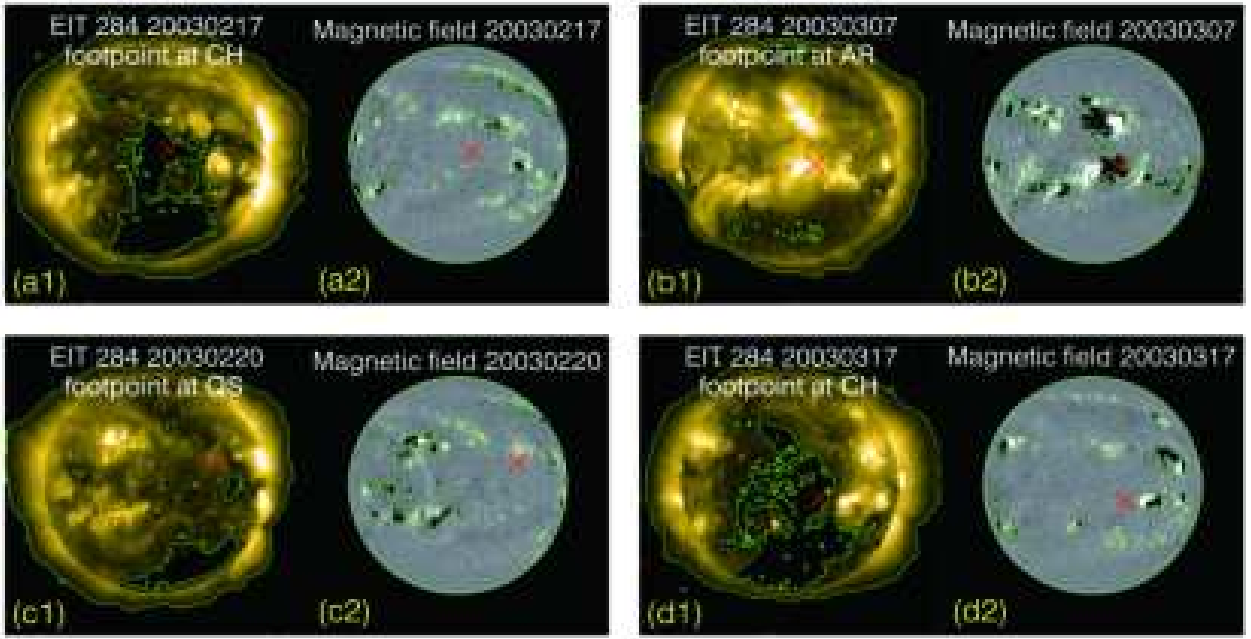}}
\caption{
 Illustration of  the classification scheme of the solar wind. EIT 284 \AA\ images (a1, b1, c1, d1) and
 corresponding photospheric magnetograms (a2, b2, c2, d2) are shown. The green contours outline the CHs and
 magnetically concentrated area boundaries. The footpoints are represented by red crosses.}
 \label{fig:Fig_eit_mdi}
 \end{center}
\end{figure*}

The categorisation scheme is demonstrated in Figure~\ref{fig:Fig_eit_mdi}.
The footpoint locations are overplotted on the
    EIT 284~\AA\ images (a1, b1, c1, d1) and MDI magnetograms (a2, b2, c2, d2)  marked by red crosses.
The classification of the source regions relies on the coronal hole
    and magnetically concentrated area boundaries.
The wind is classified as CH wind if its footpoints are
    located within the CH boundaries.
An AR wind is defined if its footpoints  fall in a magnetically concentrated area
    which is a numbered NOAA AR.
When a footpoint is positioned out of any CH and magnetically concentrated area,
    it is then marked as a QS wind.
More details on how the boundaries  are determined,
    as well as the classification of the source regions and
    the tracing back procedure, can be found in
    \citet{2015SoPh..290.1399F,2017ApJ...836..169F}.

The intervals occupied by Interplanetary Coronal Mass Ejections (ICMEs)
    were discarded where the charge states \oql\ exceed $6.008 exp(-0.00578v)$,
    where $v$ is the ICME speed in \velunit\
    \citep{2004JGRA..109.9104R}.
The daily averaged solar wind velocity was used in
    tracing the solar wind back to the source surface,
    therefore one footpoint for each day is determined.
The data used here cover the years from 2000 to 2008. This time period
   covers the solar maximum (2000 --  2001, hereafter MAX),
   the  decline (2002 -- 2006, DEC),
    and the minimum phases (2007 -- 2008, MIN) of cycle 23.
The statistical results are based on the following number of measurements.
For the full speed range (Figure~\ref{fig:space_ahe_vratio}),
    the hourly in-situ samples of the solar wind
     are 2844, 694, 1634, and 516 for the solar wind as a whole, CH, AR, and QS wind during solar maximum, respectively.
The hourly in-situ samples of the solar wind are 6362, 2340, 2282,
    and 1740, respectively during DEC,
    and 2169, 635, 352, 1182 during the MIN phase.

\section{Results and discussion}
\label{sec:results and discussion}

\subsection{Full speed range solar wind}\label{sec:full speed range}
\begin{figure*}
\begin{center}
\centerline{\includegraphics[width=1.0\textwidth]
{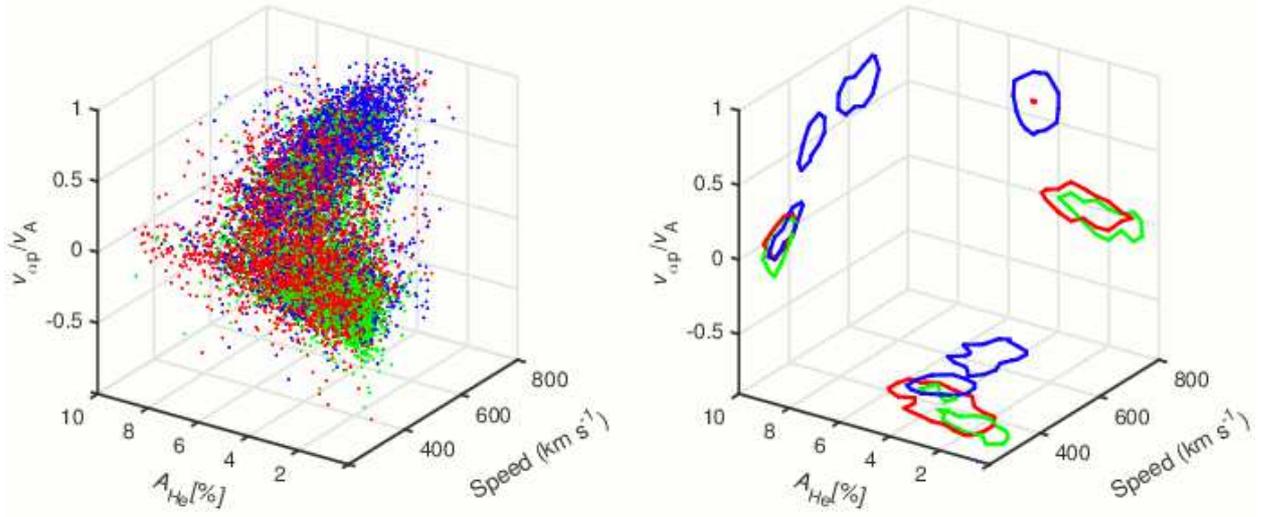}}
\caption{
 Scatter (left panel) and contour (right) plots of the solar wind in speed, \ahe, and \vratio\ space.
 Blue, red, and green represent the wind from CH, AR, and QS regions, respectively.
 The contours correspond to 50\%\ of the maximum counts for different source region solar wind.
 \label{fig:3d_vel_ahe_vratio}}
 \end{center}
\end{figure*}

Figure~\ref{fig:3d_vel_ahe_vratio}  presents the scatter (left panel) and contour (right panel) plots of the solar wind
   measurements in speed, \ahe, and \vratio\ space.
Clearly, the solar wind as a whole is separated into two main parts in the three-dimensional space
    which is more evident from the contour plots.
One part lies in the region of higher \ahe\ and \vratio, and a wider speed range, coming mainly from CHs.
In contrast, the other part has lower values of \ahe, \vratio, and a wider \ahe\ range, that mainly originates from AR and QS regions.
The quantitative analyses of \ahe\ and \vratio\ for the three source region solar wind
    are given in the following.

\begin{figure*}
\begin{center}
\centerline{\includegraphics[width=1.0\textwidth]
{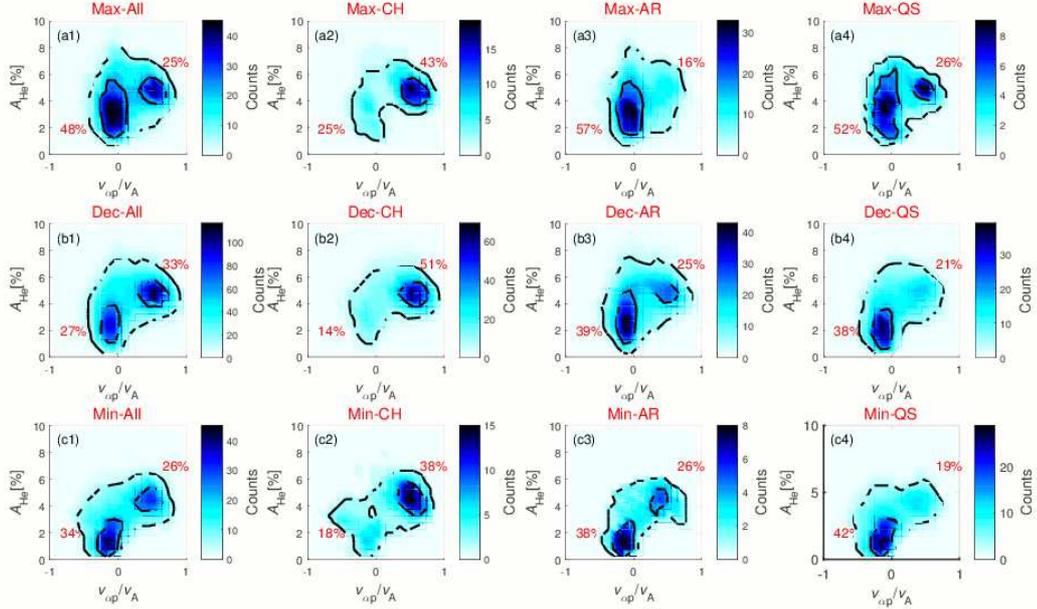}}
\caption{
 Contour plots of the solar wind originating in
    CH, AR, and QS regions in the \ahe\ and \vratio\ space.
 The first to fourth columns correspond to the wind as a whole,
    and from CH, AR, and QS regions, respectively.
The top, middle, and bottom panels represent the
    solar maximum, decline, and minimum phases.
 The percentages of the samples located in the concentrations of higher \ahe\ and \vratio\ values  (H$\_$av)
    and lower \ahe\ and \vratio\ values (L$\_$av) are also given. The solid line contour corresponds to 50\%  and the outer contour to 10\%.
 \label{fig:space_ahe_vratio}}
 \end{center}
\end{figure*}

Figure~\ref{fig:space_ahe_vratio} (first column)  shows the measurements of \ahe\ vs \vratio\
    for the solar wind as a whole.
The ranges of \ahe\ (0 -- 10) and \vratio\ ($-$1.0 -- 1.0) are divided into 20 parts.
This means that  delta \ahe\ and delta \vratio\ are 0.5 and 0.1, respectively, and the space of \ahe\ and \vratio\ is divided into 400 subsections.
We also made estimations of the \ahe\ vs \vratio\  for each individual source region,
    namely,  CH, AR and QS (second, third and fourth  columns, respectively)
    that make the total contribution in the first column of Figure~\ref{fig:space_ahe_vratio}.
 The results are also obtained for the three solar cycle phases,
    the MAX, the DEC, and the MIN as  the \ahe\ is known
    to change with the solar cycle activity
    \citep{2001GeoRL..28.2767A,2007ApJ...660..901K,2012ApJ...745..162K}.
The solar wind  as a whole (first column in Figure~\ref{fig:space_ahe_vratio}) has a  bimodal distribution in the  \ahe\ and \vratio\ space.
In order to give a quantitative evaluation of the two peaks of the distribution, we  estimated
    the proportions of the count measurements  of  \ahe\ and \vratio\  averaged over 1~hr located inside the 50\% contour lines of the solar wind as a whole.
    One of the peaks of this distribution lies in the range of higher \ahe\ and \vratio\ (hereafter H$\_$av) and has proportions  of 25\%, 33\%, and 26\%\ of the total counts for the solar MAX, DEC and  MIN.   The proportions are given in each panel of Figure~\ref{fig:space_ahe_vratio}.
In  H$\_$av the \ahe\ ranges are  3.75 -- 5.75, 3.75 -- 5.75, and 3.5 -- 5.25, and
     the \vap\ ranges are 0.3 -- 0.7, 0.3 -- 0.8, and 0.2 -- 0.7, respectively.
In contrast, the other peak covers lower values of \ahe\ and \vratio\  with  contributions of
    48\%, 27\%, and 34\% during the MAX, DEC, and MIN (hereafter L$\_$av).
The corresponding \ahe\ ranges are  1.25 -- 5.50, 1.00 -- 3.75, and 0.25 -- 3.00, while the
 \vap\ ranges are $-$0.3 -- 0.2, $-$0.2 -- 0.1, and $-$0.2 -- 0.1.
It is notable that
     the \ahe\ distribution ranges of H$\_$av are narrower than
     the ranges of L$\_$av, 2.0 vs 4.25, 2.0 vs 2.75, and 1.75 vs 2.75,
     during the MAX,  DEC and MIN, respectively.
We note that \citet{2011ApJ...728L...3B} have also obtained a bimodal distribution in the
    \ahe\ and \vratio\ space.
However, the difference from their study (see their Figure~3) is that we give the distributions for
    the solar wind originating from various source regions.

We then estimated  the count contributions  of  H$\_$av  and L$\_$av
    for each of the solar wind source regions
    by applying the same \ahe\ and \vratio\ ranges
    as estimated from the whole Sun bimodal distribution.
A noticeable  finding here is that
    the CH wind counts are concentrated  in the H$\_$av
    (Figure~\ref{fig:space_ahe_vratio}, second column).
The lower limit contributions of  H$\_$av are much higher
   than those of L$\_$av for CH wind.
In contrast, for the AR and QS wind the proportions of L$\_$av  are much higher than
    H$\_$av (Figure~\ref{fig:space_ahe_vratio}, third and fourth column).
Thus, the AR and the QS wind counts are predominantly located in the L$\_$av.

How the results change if the three types (CH, AR, and QS) solar wind
    are determined more restrictively?
Usually the footpoints of the solar wind stay in a particular region
    for a few days (as shown in Figure~1 of \citet{2017ApJ...836..169F}).
A boundary wind is defined when the footpoint of the solar wind  moves from
    one region to another  \citep{2002JGRA..107.1488N}.
Here, if the footpoint is located at the same region for more than 2 days,
    the solar wind associated with the first and last 12 hours,
    between the change of the footpoint location is defined as a boundary wind.
This way only the core wind for a certain region can be extracted.
As expected, the samples of the data become smaller.
However, the distribution characteristics of the core solar wind are
    almost the same as in the original selection scheme.
The only clear difference is that the proportions of
     L$\_$av decrease for the CH wind.

Several studies indicate that the solar wind produced by the WTD mechanism
 has  higher \ahe\ and \vap,
    while the solar wind may have lower \ahe\ and \vap\ when the RLO mechanism is at work.
While it is challenging to obtain direct observations of the \ahe\  in the solar corona,
first ionisation potential (FIP)  bias measurements are more readily available.
It is found that generally the FIP bias is higher in AR and QS regions
    (mainly occupied by closed loops)
     than in CH regions (generally taken up by open magnetic field lines)
     \citep{2001ApJ...555..426W, 2005JGRA..110.7109F,
     2011ApJ...727L..13B, 2013ApJ...778...69B}.
It is believed that the reason for the enrichment of
    the low FIP ions
    in the corona and the solar wind is that they are ionized earlier
    in comparison to high FIP elements.
The helium has the highest FIP and remains neutral longest.
This results in the enrichment/depletion of low FIP elements/helium
    because only ions interact with waves
    \citep{2012ApJ...744..115L, 2015LRSP...12....2L, 2017ApJ...844..153L}.
It means that the helium abundance should be inversely proportional to
    the low FIP bias elements in the corona.
Thus,  the \ahe\ is higher in open magnetic field structures
 and
    lower in closed loops
   if the above mechanism is valid.
The helium abundance for the fast SW coming from large CHs
    is higher and remarkably stable
    \citep{2006SSRv..124...51S}.
In contrast, \citet{2012ApJ...754...65R}
    suggested that the helium is depleted in closed loops and
    the depletion efficiency is higher in larger loops,
    and lower in smaller loops.
Furthermore, from  simulations,
    \citet{2017ApJ...844..153L} showed that
    the \ahe\ is higher in open magnetic field regions
    and it is lower in closed loops.
\citet{2009JGRA..114.4103S}
    found that the solar wind that comes from big streamers has lower helium abundance.
The solar wind can be produced by interchange reconnection in the streamers
    \citep{2016JGRA..121...19H}.
This gives the observational support to the notion that the \ahe\
    is lower in the closed loops
    as streamer structures are composed by very large closed loops.

As already mentioned, the speeds of helium ions are usually larger than protons,
    although the helium ions are heavier than protons.
Generally, it is believed that the helium ions are heated by resonant
    wave-particle interactions
    \citep[e.g.,][]{1978JGR....83...97H,1982Ap&SS..81..295M,1984JGR....89.6613I}
    with waves preferentially heating the heavy ions,
    making them faster than protons.
This means that the wave acceleration is the reason why
    there is a speed difference between helium ions and protons
    \citep{2008ApJ...678.1480C}.
The fact that \vap\ decreases with heliocentric distances
    \citep{1982JGR....87...35M,1996JGR...10117047N,2001JGR...106.5693R}
    suggests that the speed difference between the helium ions and protons
 is produced near the Sun.
Therefore, it supports the idea that the wave super-acceleration of helium ions
    takes place near the Sun,
    possibly in the region of acceleration of the solar wind
    \citep{1996JGR...10117047N}.
For the solar wind that escapes directly along open magnetic field lines,
    the wave accelerated solar wind starts from the chromosphere, whereas the solar wind released from closed loops is
    initiated higher in the solar atmosphere.
It is possible that
    the solar wind that escapes directly along open magnetic field lines
    has more time to make a speed difference compared with the wind
    released from loops. This speculation, however, needs to be further supported by modeling.
The above effects indicate that the wind that escapes directly along
    open magnetic field lines (treated by the WTD models) has larger \vap,
    whereas the solar wind released from closed loops (e.g., RLO models)
    has smaller \vap\
    \citep{2016ApJ...829..117S}.

However, the \vap\ will also be affected in the following propagation process.
First, it is constrained by various instabilities such as the
    Alfv\'en/ion-cyclotron and fast-magneto\-sonic/whistler instabilities
    \citep{2013ApJ...777L...3B}.
Qualitatively, those instabilities are only valid
    when the \vap\ nears or exceeds the
    local Alfv\'en speed
    \citep{2000JGR...105.7483L, 2013ApJ...764...88V, 2009PhPl...16d2901L},
    therefore they have smaller effect on the solar wind with small \vratio.
Second, collisions can reduce \vap\
    in the solar wind  \citep[e.g.,][]{1982JGR....87...52M, 2008PhRvL.101z1103K}.
\citet{2011ApJ...728L...3B} found a general inverse relationship
    between the \vratio\ and collisional age,
    which means the collisions are important.
This should be more significant for the slow SW which has higher collisional age.
However, the distribution range in collisional age and \vratio\ space are very large
    (Figure~2 (b) in \citeauthor{2011ApJ...728L...3B},\citeyear{2011ApJ...728L...3B})
which indicates that \vap\ may still persist in part of the slow speed SW
    (speed less than 500 \velunit).
Here, the distributions in \ahe\ and \vratio\ space for the intermediate SW (with speed greater than 400 \velunit\ and less than 500 \velunit)
    demonstrate that a higher \vratio\  at $\sim$0.4 could still be found in the solar wind whose
    speed is less than 500 \velunit\
     (see Section~\ref{sec:fast and slow solar wind} and Figure~\ref{fig:space_ahe_vratio_MSW} below for justifications).

 \citet{2017ApJ...836..169F} suggested that
    the two-peak distribution of CH wind and
    the anti-correlation between the speed and  \oql\
    can be explained qualitatively by both the  WTD and RLO models,
    implying that the combination of the two classes of mechanisms may be at work
    \citep{2009LRSP....6....3C}.
The clear bimodal distribution in space of
    \ahe\ and \vratio\ for the solar wind as a whole
    (Figure~\ref{fig:3d_vel_ahe_vratio} and the first column of Figure~\ref{fig:space_ahe_vratio})  can also  result from the interplay of the direct plasma release mechanism along open magnetic field lines treated by the WTD models and via reconnection  between closed and open fields in the RLO models.
This is consistent with the idea that the WTD and RLO scenarios do not need to be `mutually exclusive with each other' as suggested by \citet{2009LRSP....6....3C} and \citet{2016SSRv..201...55A}.

The distributions in space of \ahe\ and \vratio\ for the solar wind that originates from
   various source regions (the second to fourth columns of Figure~\ref{fig:space_ahe_vratio})
    provide additional support to the notion that  possibly the WTD and RLO mechanisms work together.
Spectroscopic observations have shown that there exists a stable outflow
    at the base of CH regions
    \citep{1999Sci...283..810H,2003A&A...399L...5X,
    2004A&A...424.1025X,2005Sci...308..519T}
    and the outflow usually corresponds to concentrations
    of  unipolar magnetic fields
    \citep{2003A&A...399L...5X,2004A&A...424.1025X}
    where the open magnetic field lines are rooted.
On the other hand, closed loops dominate in ARs and QS,
    and therefore, the solar wind plasma is more likely to be
    released by magnetic reconnection between
    closed loops and open magnetic field lines
    \citep{2002JGRA..107.1488N,2005JGRA..110.7109F, 2008ApJ...676L.147H,
    2010A&A...516A..14H, 2012A&A...545A...8Z, 2012SoPh..281..237V,2014SoPh..289.3799C,
    2014SoPh..289.4151M, 2017SoPh..292...46B}.

The  low count number at L$\_$av and H$\_$av for the
    three types solar wind can also be explained reasonably by the above suggestion.
For the CH wind  (Figure~\ref{fig:space_ahe_vratio}, second column)
    these  L$\_$av  measurements could be interpreted
    as related to magnetic reconnection between open magnetic field lines and loops along CH boundaries, as well as
 loops associated with small magnetic bipoles,
    in particular with the emergence of ephemeral regions.
In ARs  (Figure~\ref{fig:space_ahe_vratio}, third column),
    a low count H$\_$av is possibly related to small-scale coronal holes
    adjacent to ARs which are
    often obscured by overlying large loops coming out of the ARs
    \citep[e.g.,][]{2017ApJ...841...94W}.
The domination of L$\_$av in the QS is logical
    as this region is predominately seeded by closed loops.
The existence of open field lines in QS regions explains the presence of H$\_$av counts that
   are smaller but still significant
    \citep{1997GeoRL..24.1159W, 1997ApJ...489L.103H, 2000JGR...10512667W}.

The observational results for the \ahe\ distribution ranges of  H$\_$av and L$\_$av
    in Figure~\ref{fig:space_ahe_vratio}
    can be reasonably interpreted as an interplay of
     the WTD and RLO  mechanisms as suggested above.
The \ahe\ distribution ranges
    are smaller for the H$\_$av (2.00, 2.00, and 1.75 during MAX, DEC, and MIN)
    compared with the L$\_$av (4.25, 2.75, and 2.75).
The simulations by
    \citet{2012ApJ...754...65R}
    have shown that closed loops show a depletion effect on helium ions,
    and the depletion efficiency is higher in larger loops
    and lower in smaller loops.
Therefore, the distributions of \ahe\ are shifted towards lower values
    and the distribution ranges are wider for the L$\_$av
for regions populated by larger loops
    \citep[e.g.,][]{2004SoPh..225..227W,2005JGRA..110.7109F}.

\subsection{Slow, intermediate, and fast solar wind}\label{sec:fast and slow solar wind}

\begin{figure*}
\begin{center}
\centerline{\includegraphics[width=1.0\textwidth]
{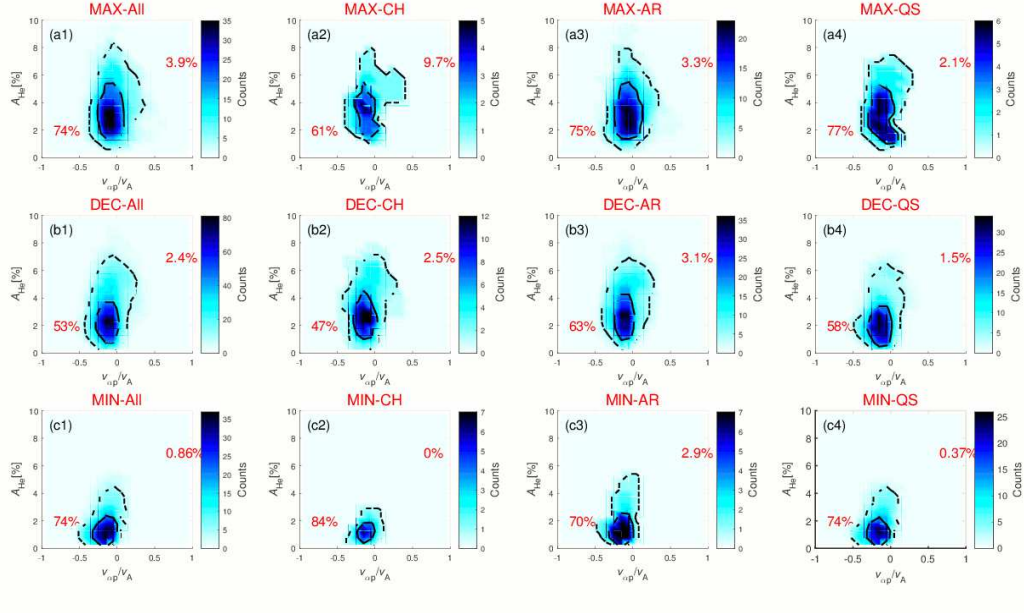}}
\caption{
As in Figure~\ref{fig:space_ahe_vratio},
    but for the slow SW only.
 \label{fig:space_ahe_vratio_SSW}}
 \end{center}
\end{figure*}

\begin{figure*}
\begin{center}

\centerline{\includegraphics[width=1.0\textwidth]
{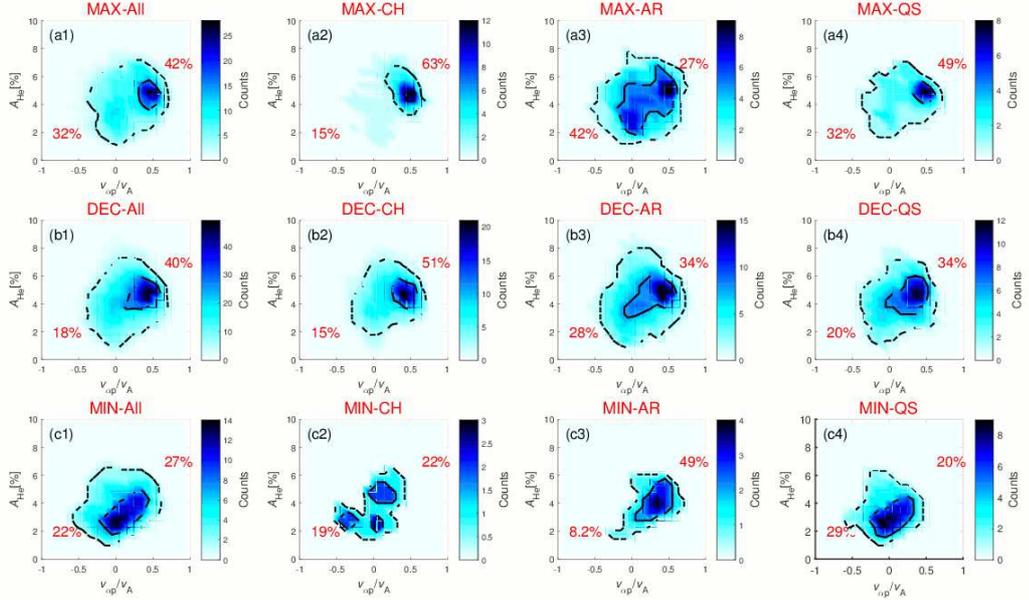}}
\caption{
As in Figure~\ref{fig:space_ahe_vratio},
    but for the intermediate SW only.
 \label{fig:space_ahe_vratio_MSW}}
 \end{center}
\end{figure*}

\begin{figure*}
\begin{center}
\centerline{\includegraphics[width=1.0\textwidth]
{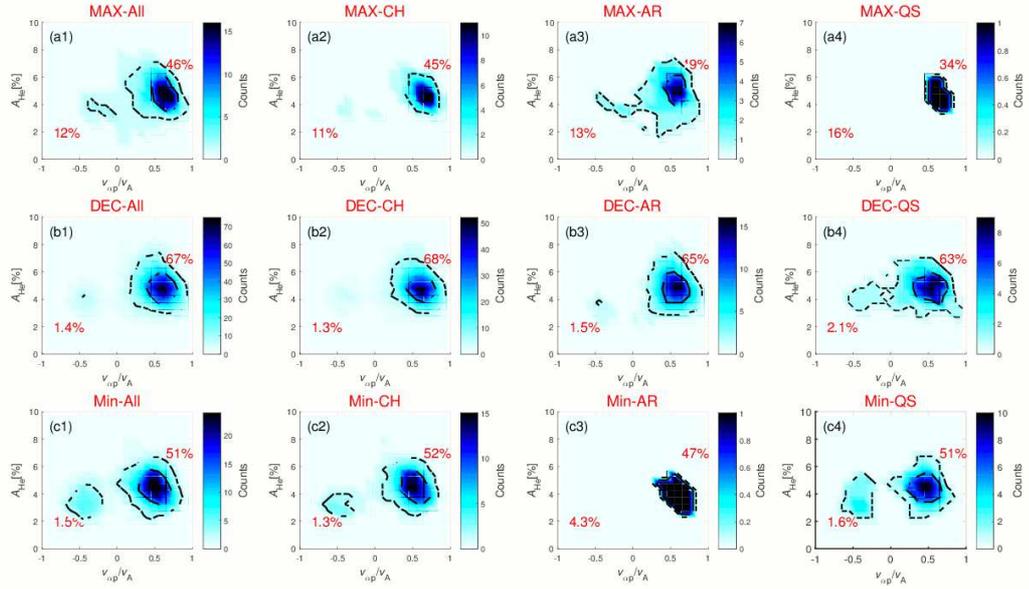}}
\caption{
As in Figure~\ref{fig:space_ahe_vratio},
    but for the fast SW only.
 \label{fig:space_ahe_vratio_FSW}}
 \end{center}
\end{figure*}

The results for the solar wind in full speed range
    (Figure~\ref{fig:space_ahe_vratio}),
    show a clear bimodal distribution denoted as H$\_$av and L$\_$av.
Generally, the solar wind is divided into two categories, fast and slow SW, and
 two thresholds are usually chosen, 400 \velunit\
    \citep[e.g.][]{2006SSRv..124...51S}  or 500 \velunit\
    \citep[e.g.][]{2015SoPh..290.1399F,2016ApJ...829..117S,2017ApJ...836..169F}.
To separate the distribution characteristics of Figure~\ref{fig:3d_vel_ahe_vratio},
    we divided the solar wind into three categories,
    slow SW (speed of less than 400 \velunit),
    intermediate SW (speed greater than 400 \velunit\ and less than 500 \velunit),
    and fast SW (greater than 500 \velunit).
To explore the distribution characteristics of the solar wind
    in different speed ranges
    for the three source regions,
    we produced  distributions in \ahe\ and \vratio\ space for the
    slow SW, intermediate SW,  and fast SW given in Figure~\ref{fig:space_ahe_vratio_SSW},
    Figure~\ref{fig:space_ahe_vratio_MSW},
    and  Figure~\ref{fig:space_ahe_vratio_FSW}, respectively.

Figure~\ref{fig:space_ahe_vratio_SSW}
    shows the distributions in \ahe\ and \vratio\ space for the slow SW only.
The proportions of L$\_$av are much higher than that of H$\_$av
    (generally far below 10\%, except for CHs).
The measurements for the slow SW are mainly concentrated at lower
    \ahe\ and \vratio\ values with wider \ahe\ distribution ranges
    for all three types slow SW.
In contrast, the fast SW (Figure~\ref{fig:space_ahe_vratio_FSW})
    is  mainly distributed at higher \ahe\ and \vratio\ values with
    narrower \ahe\ distribution ranges for all three types fast SW.
The difference in the proportions of H$\_$av and L$\_$av for the intermediate SW
   are  obvious comparing the slow SW (dominant L$\_$av) and fast SW (dominant H$\_$av).
 H$\_$av and L$\_$av are both present in all three types intermediate SW  (Figure~\ref{fig:space_ahe_vratio_MSW}), with  H$\_$av that is dominant during the MAX and DEC phases for the whole SW.
In general, the distributions for the intermediate SW are more complex than for the slow SW and fast SW which could be seen in Figure~\ref{fig:space_ahe_vratio_MSW}, second (CHs), third (ARs), and fourth column (QS).
As we suggested, the L$\_$av is produced by the RLO mechanism,
    whereas the H$\_$av associates with the WTD mechanism.
Our results indicate that the slow SW
    is mainly produced by the RLO mechanism,
    in contrast the fast SW  is
    mainly generated by the WTD regardless of the source types of the solar wind.
The WTD and RLO mechanisms are both present in the intermediate SW, with
a different input of each mechanism for the different source regions and during different phases of the solar cycle activity.
   The L$\_$av for the AR region solar wind is most probably related to AR edge outflows (see Sect.~\ref{sect_intro}) released into the solar wind through magnetic reconnection of the AR loops and open magnetic field lines of adjacent CHs.

Based on in-situ observations,
    \citet{2016ApJ...829..117S} suggested that the solar wind
with a  speed of  less than 500 \velunit\ can be divided into two types,
    one that is alike the fast SW and the other originating from closed loops.
It is clear that the fast SW and ``the wind originating from closed loops'' in \citet{2016ApJ...829..117S}
   can be associated with both the WTD and RLO mechanisms.
Based on the present statistical results, an interplay of the WTD and RLO mechanisms
    is present for the solar wind of less than 500~\kms.
Therefore, our results give the observational support to the suggestion  of
    \citet{2016ApJ...829..117S}.

\section{Summary and concluding remarks} \label{sec:conclusion}
The main purpose of the present work was to examine
    the statistical properties of the \ahe\ and \vratio\
    and their distributional characteristics in space of \ahe\ and \vratio\
    for three source region solar wind, CHs, ARs and QS.
The main results are summarised as follows:

\begin{enumerate}
\item
{ We found bimodal distributions of the solar wind
    as a whole in the \ahe\ and \vratio\ space.
One peak lies in the range of higher values of \ahe\ and \vratio.
In contrast, the other peak is located at lower values of \ahe\ and \vratio.
The analysis for the three source region solar wind shows that
     the CH wind counts are concentrated at higher \ahe\ and \vratio\ values
      with narrower \ahe\ distribution ranges,
     while the AR and QS wind is mainly located at lower \ahe\ and \vratio\
      with larger  \ahe\ distribution ranges}.

\item
{ Almost all of the slow SW (fast SW) measurements are concentrated at
    lower (higher) \ahe\ and \vratio\ values with wider (narrower) \ahe\ distribution ranges
    regardless of the source region type.
In contrast, the H$\_$av and L$\_$av are both present in all three types
    (CH, AR, and QS) intermediate SW.}

\end{enumerate}

The results demonstrate that there are clear differences of \ahe\ and \vratio\
    for the three source region solar wind.
This indicates that the configuration of the magnetic field
    has influence on the \ahe\ and \vratio\ properties.
We suggest that the two solar wind generation mechanisms,
    the wave-turbulence-driven (WTD)
   and the reconnection-loop opening (RLO),  work in parallel
   in all solar wind source regions.
In CH regions WTD plays a major role,
    whereas RLO is more important in  AR and QS regions.

The statistical results for different speed range solar wind indicate that
    the slow SW (speed less than 400 \velunit) is mainly produced by the RLO mechanism,
    in contrast the fast SW (speed greater than 500 \velunit) is mainly generated by the WTD mechanism
    regardless of the source types of the solar wind.
Whereas both the WTD and RLO mechanisms play role for the
    generation of the intermediate SW
    (speed range  400 -- 500 \velunit).

The future Solar Orbiter mission that
    comes as close as 0.285 AU to the Sun
    should help reduce the uncertainties in tracing the solar wind back to the Sun and thus bring  more accurate evaluation of the properties of  \ahe\ and \vap\ for different source region solar wind.

\section*{Acknowledgements}
The authors thank the referee Pascal Demoulin for the very helpful
comments and suggestions.
Analysis of Wind SWE observations is supported by NASA
    grant NNX09AU35G.
SoHO is a project of international cooperation
    between ESA and NASA.
This research is supported by
    the National Natural Science Foundation of China (41604147,
    41474150 and 41474149).
H.F. thanks the Shandong provincial Natural Science Foundation (ZR2016DQ10).
Z.H. thanks Young Scholars Program of Shandong University, Weihai.

\bibliographystyle{mnras}
\bibliography{reference_Helium}

\bsp    
\label{lastpage}
\end{document}